\documentclass[12pt]{iopart}
\expandafter\let\csname equation*\endcsname\relax
\expandafter\let\csname endequation*\endcsname\relax 
\usepackage{iopams}
\usepackage{epsfig}
\usepackage{amssymb,amsmath,amsfonts,dsfont}
\usepackage{graphicx}
\usepackage{float}
\usepackage{lipsum}
\usepackage{array}
\usepackage{color}
\usepackage[numbers,sort&compress,square]{natbib}
\begin{document}
\title{Parallel quantum trajectories via forking for sampling without redundancy}
\author{Daniel K. Park$^{1,2,*}$, Ilya Sinayskiy$^{3,4,\dagger}$, Mark Fingerhuth$^{3,5,\ddagger}$, Francesco Petruccione$^{1,3,4,\S}$ and June-Koo Kevin Rhee$^{1,2,3,\P}$}
\address{$^1$School of Electrical Engineering, KAIST, Daejeon, 34141, Republic of Korea}
\address{$^2$ITRC of Quantum Computing for AI, KAIST, Daejeon, 34141, Republic of Korea}
\address{$^3$Quantum Research Group, School of Chemistry and Physics, University of KwaZulu-Natal, Durban, KwaZulu-Natal, 4001, South Africa}
\address{$^4$National Institute for Theoretical Physics (NITheP), KwaZulu-Natal, 4001, South Africa}
\address{$^5$ProteinQure Inc., Toronto, M5T 2C2, Canada}
\ead{$^*$kpark10@kaist.ac.kr, $^\dagger$sinayskiy@ukzn.ac.za, $^\ddagger$markfingerhuth@gmail.com, $^\S$petruccione@ukzn.ac.za, $^\P$rhee.jk@kaist.edu}
\begin{abstract}
The computational cost of preparing a quantum state can be substantial depending on the structure of data to be encoded. Many quantum algorithms require repeated sampling to find the answer, mandating reconstruction of the same input state for every execution of an algorithm. Thus, the advantage of quantum computation can diminish due to redundant state initialization. We present a framework based on quantum forking that bypasses this fundamental issue and expedites a family of tasks that require sampling from independent quantum processes. 
Quantum forking propagates an input state to multiple quantum trajectories in superposition, and a weighted power sum of individual results from each trajectories is obtained in one measurement via quantum interference.
The significance of our work is demonstrated via applications to implementing non-unitary quantum channels, studying entanglement and benchmarking quantum control. A proof-of-principle experiment is implemented on the IBM and Rigetti quantum cloud platforms.
\end{abstract}
\maketitle
\def\one{{\mathchoice {\rm 1\mskip-4mu l} {\rm 1\mskip-4mu l} {\rm \mskip-4.5mu l} {\rm 1\mskip-5mu l}}}

\section{Introduction}
\label{sec:intro}
Designing an efficient quantum algorithm to solve a computational task does not alone ensure a quantum advantage over a classical counterpart, but there must also be an efficient procedure to prepare the desired initial quantum state. Existing methods for preparing an arbitrary quantum state~\cite{PhysRevA.97.052329,PhysRevA.64.014303,PhysRevA.73.012307,Mosca:01,2002quant.ph.8112G,PhysRevA.83.032302,PhysRevLett.92.177902,VENTURA2000273,Mottonen:2005:TQS:2011670.2011675}, a famous example being quantum random access memory (QRAM)~\cite{QRAMPhysRevLett.100.160501,QRAMPhysRevA.78.052310,PhysRevA.86.010306,1367-2630-17-12-123010,FFQRAM-SciRep-Park}, introduce resource overheads, even though the hardware and process complexities may scale efficiently with respect to the size of the data to be encoded. It is therefore imperative to minimize the number of state preparation routines. An input quantum superposition state cannot be reused for another task once measured due to the measurement postulate of quantum mechanics. Moreover, the quantum state cannot be cloned. Hence, in general, one is forced to generate an input state in every execution of a quantum algorithm. However, many quantum information processing (QIP) tasks rely on repeating the measurement for sampling the answer. Thus the true advantage of harnessing quantum mechanics for information processing becomes unclear when the aforementioned redundancy is imposed. As a means to circumvent this fundamental issue in certain applications, quantum forking (QF) was introduced in~\cite{FFQRAM-SciRep-Park}, motivated by forking in classical operating systems that creates a separate address space for a child process for multitasking~\cite{UnixForking}. In~\cite{FFQRAM-SciRep-Park}, the application was limited to estimating the inner product of quantum states.

In this work, we present quantum forking-based sampling (QFS) to accelerate various tasks that require adding the results from independent quantum trajectories as a convex combination. With this framework, the number of state preparation routines and measurements required for performing a weighted power summation of measurement outcomes sampled from an arbitrary number of independent quantum processes remains constant, though this is at the cost of introducing a control qudit, ancilla qubits in arbitrary states, and a series of controlled swap gates. Moreover, QFS can be used to efficiently measure an arbitrary observable since a convex combination of Hermitian operators is also Hermitian. This is particularly useful when directly measurable observables are limited by experimental constraints. The weighted power summation of quantities measured in multiple quantum processes is required in solving various problems in quantum science as we demonstrate with examples in implementing a convex combination of quantum channels, detecting entanglement, and characterizing quantum control. We realize a proof-of-principle experiment of a QFS on two quantum computers in the cloud, the IBM Q 5 Tenerife~\cite{tenerife} and Rigetti 16Q Aspen-1~\cite{aspen}, demonstrating the feasibility of the technique with near-term quantum hardware. QFS is fundamentally intriguing as it shows that without violating the no-cloning theorem, a single quantum state with ancillary space suffices to accomplish tasks that would naively require many copies of the quantum state.

Furthermore, when large-scale quantum hardware becomes available, a quantum operating system that deploys efficient resource management and acts as an interface between qubits and quantum programs is of fundamental importance~\cite{QOS}. Along with QRAM, the quantum forking algorithm developed here has the potential to be used as a building block for such a quantum operating system.

The remainder of the paper is organized as follows. Section~\ref{sec:qf} reviews and further generalizes the quantum forking framework which was first introduced in~\cite{FFQRAM-SciRep-Park}. Next, we describe the QFS protocol for an arbitrary weighted power summation of results from $d$ independent quantum trajectories using a constant number of state preparations and measurements in section~\ref{sec:ExpVal} and section~\ref{sec:Projective} for the expectation value measurement and the projective measurement, respectively. In section~\ref{sec:app}, we illustrate how the number of state preparations can be reduced in constructing a convex combination of quantum channels, which naturally enables the quantum channel twirling, detecting entanglement witnesses, and purity benchmarking as example applications for which QFS can be useful. Section~\ref{sec:exp} demonstrates the proof-of-principle implementation of a simple QFS experiment on the cloud quantum computers, and section~\ref{sec:conclusion} concludes.
\section{Quantum forking}
\label{sec:qf}
\begin{figure}[t]
\centering
\includegraphics[width=0.4\linewidth]{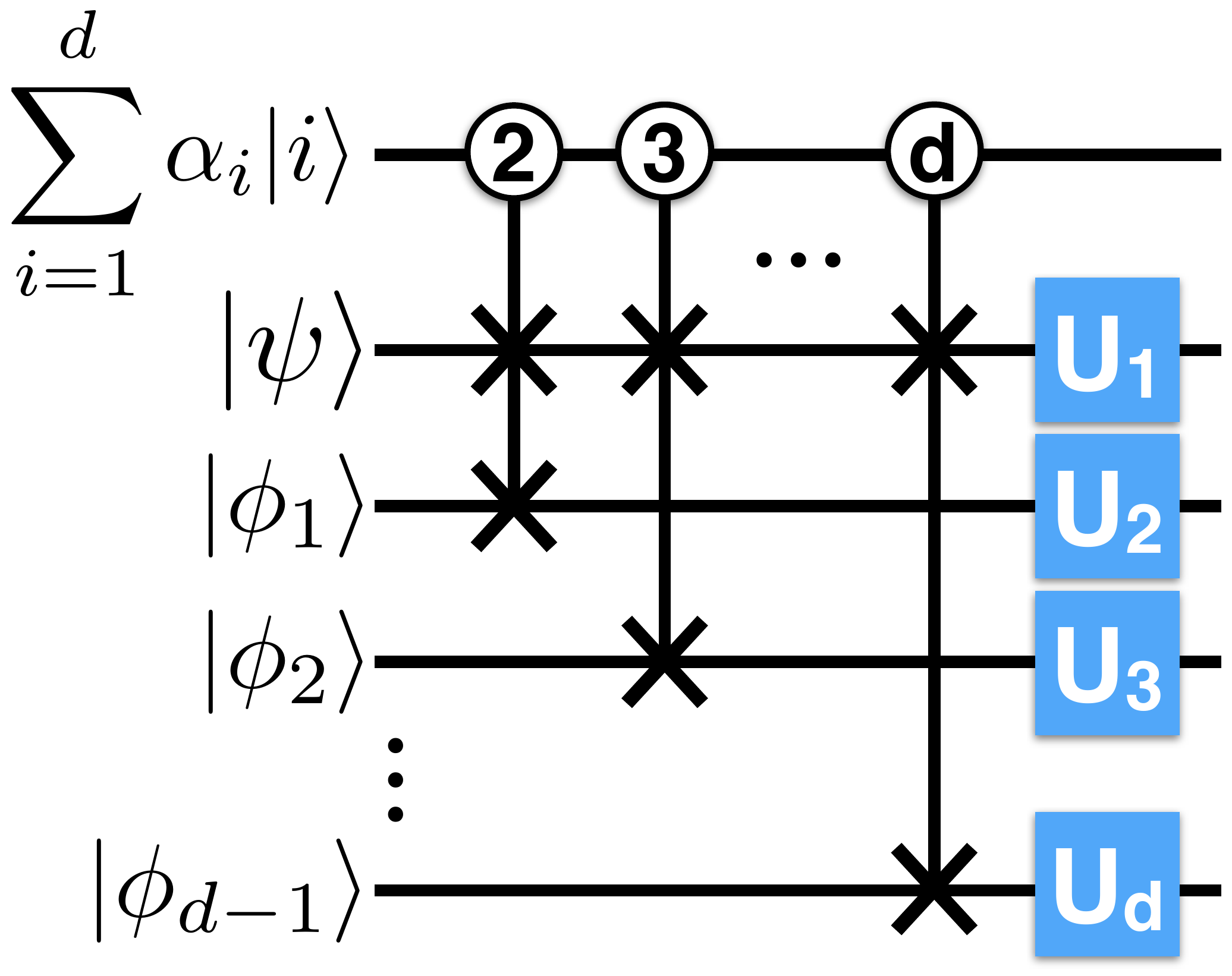}
\caption{Quantum forking circuit to create $d$ independent processes, denoted by $U_i$, in superposition. The swap operation is represented by two $\times$ symbols connected with a vertical line. The numbers in circles indicate the qudit state that activates the controlled swap gate.}
\label{fig:1}
\end{figure}

Quantum forking is a process that creates an entangled state to allow a target quantum state $|\psi\rangle$ to undergo $d$ independent evolutions in superposition~\cite{FFQRAM-SciRep-Park}. At the end of the computation, some measurement procedure is employed to find the desired answer more efficiently compared to when the quantum state undergoes these evolutions one at a time. This can be achieved by coupling the target state $|\psi\rangle$ to a control qudit (or $\log_2(d)$ qubits) and $d-1$ ancilla qubits in some arbitrary state $|\phi_j\rangle$, such that the target state undergoes independent processes in $d$ orthogonal subspaces. The quantum circuit for realizing QF is depicted in figure~\ref{fig:1}.
The quantum forking is initiated by the series of controlled swap (c-\texttt{swap}) gates controlled by the state of the qudit. These gates evolve the total state as
\begin{align}
\label{eq:1}
|\Psi\rangle=&\alpha_1|1\rangle\otimes|\psi\rangle\otimes|\phi_1\rangle \otimes|\phi_2\rangle\otimes\ldots \otimes|\phi_{d-1}\rangle\nonumber \\
+&\alpha_2|2\rangle \otimes|\phi_1\rangle \otimes|\psi\rangle \otimes|\phi_2\rangle\otimes\ldots\otimes|\phi_{d-1}\rangle\nonumber \\
+&\alpha_3|3\rangle \otimes|\phi_2\rangle \otimes|\phi_1\rangle \otimes|\psi\rangle\otimes\ldots\otimes |\phi_{d-1}\rangle\nonumber \\
+&\ldots\nonumber \\
+&\alpha_d|d\rangle \otimes|\phi_{d-1}\rangle \otimes|\phi_1\rangle \otimes|\phi_2\rangle\otimes\ldots \otimes|\psi\rangle.
\end{align}
Hereinafter, the tensor product symbol is omitted for brevity when the meaning is clear. Equation~(\ref{eq:1}) shows that QF creates an entangled state whereby the target quantum state $|\psi\rangle$ is encoded in a different qubit for each subspace referenced by the control qudit. Thus, by applying local unitaries, the target quantum state can undergo $d$ independent processes simultaneously. For instance, the total state after the application of local unitary operators becomes
\begin{align}
|\Psi\rangle=&\alpha_1|1\rangle U_1|\psi\rangle U_2|\phi_1\rangle U_3|\phi_2\rangle\ldots U_d|\phi_{d-1}\rangle\nonumber \\
+&\alpha_2|2\rangle U_1|\phi_1\rangle U_2|\psi\rangle U_3|\phi_2\rangle\ldots U_d|\phi_{d-1}\rangle\nonumber \\
+&\alpha_3|3\rangle U_1|\phi_2\rangle U_2|\phi_1\rangle U_3|\psi\rangle\ldots U_d|\phi_{d-1}\rangle\nonumber \\
+&\ldots\nonumber \\
+&\alpha_d|d\rangle U_1|\phi_{d-1}\rangle U_2|\phi_1\rangle U_3|\phi_2\rangle\ldots U_d|\psi\rangle.
\end{align}
The ancilla qubits can be untouched if desired, by using controlled unitary operators. 

\section{Quantum forking for sampling}
\label{sec:qfs}
\subsection{Expectation value measurement}
\label{sec:ExpVal}
Quantum forking can be furnished with a measurement procedure to evaluate a weighted power sum of the following form with only a constant number of initial state generations:
\begin{equation}
\label{eq:3}
    \sum_{i=1}^{d}p_i\prod_{j=1}^q\langle M_{i,j}\rangle,
\end{equation}
where $p_i$ is a non-negative real number satisfying $\sum p_i=1$, $q$ is a positive integer, and $M_{i,j}$ is an observable. In general, estimating equation~(\ref{eq:3}) to within $\epsilon$ with a probability of error $\delta$ requires $O(q'd\log(1/\delta)/\epsilon^2)$ state preparations~\cite{estimate}, where $q'\le q$ is the number of unique observables in the non-linear sum. QFS yields the same result with $O(q\log(1/\delta)/\epsilon^2)$ state preparations, reducing the time complexity by about a factor of $d$. In the following, we explain QFS for linear ($q=1$) and quadratic ($q=2$) sums using the cases where $M_{i,1}=M_{i,2}$ and $d=2$.

\subsubsection{Linear summation}
\begin{figure}[t]
\centering
\includegraphics[width=0.7\linewidth]{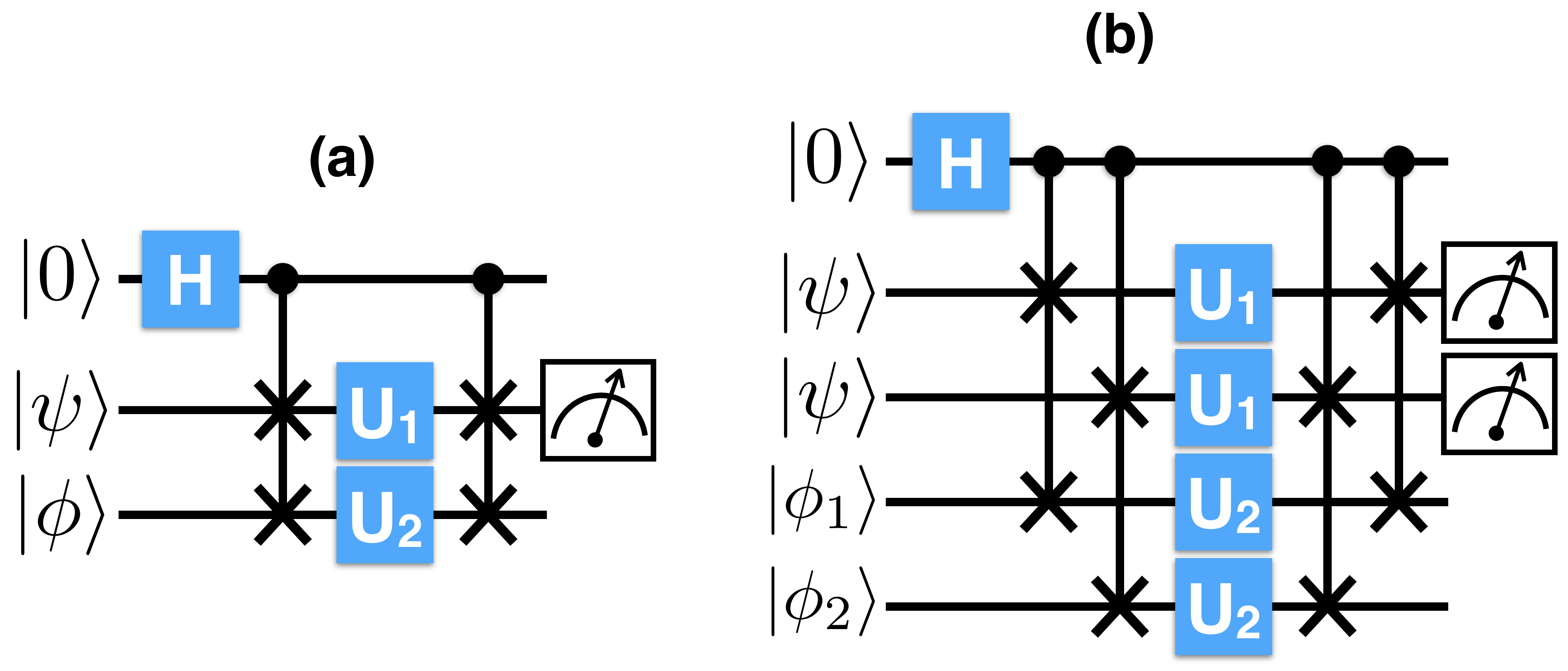}
\caption{Quantum circuits with forking for (a) linear and (b) quadratic summations of two measurement outcomes. The initial state of the target qubit is denoted as $|\psi\rangle$, and $|\phi\rangle$ and $|\phi_{1,2}\rangle$ represent an arbitrary initial state of the ancilla qubits. The target qubit undergoes two independent processes, $U_1$ and $U_2$, in superposition via quantum forking. Performing the expectation value (projective) measurement on the target qubit provides (a) linear or (b) quadratic sum of expectation values (probabilities) as described in section~\ref{sec:ExpVal} (section~\ref{sec:Projective}).}
\label{fig:2}
\end{figure}

Adding two expectation value measurement outcomes with equal weights can be done with one control qubit and one ancilla qubit in an arbitrary state $|\phi\rangle$, as depicted in figure~\ref{fig:2}(a). First, the control qubit is prepared in the equal superposition state via the Hadamard gate (denoted H). The first c-\texttt{swap} gate then yields
\begin{equation}
\label{eq:4}
    |\Phi_1\rangle=\frac{|0\rangle|\psi\rangle|\phi\rangle+ |1\rangle|\phi\rangle|\psi\rangle}{\sqrt{2}}.
\end{equation}
The two local unitaries transform the above state to
\begin{equation}
\label{eq:5}
    |\Phi_2\rangle=\frac{|0\rangle U_1|\psi\rangle U_2|\phi\rangle+ |1\rangle U_1|\phi\rangle U_2|\psi\rangle}{\sqrt{2}}.
\end{equation}
Finally, another c-\texttt{swap} gate \textit{unforks}, i.e., reverses the forking, such that the final state is
\begin{equation}
    |\Phi_f\rangle=\frac{|0\rangle U_1|\psi\rangle U_2|\phi\rangle+
    |1\rangle U_2|\psi\rangle U_1|\phi\rangle}{\sqrt{2}}.
\end{equation}
Now the expectation value of an observable $M$ measured on the target qubit gives the desired average of the two quantities:
\begin{align}
\label{eq:7}
    \langle M \rangle&=\langle \Phi_f| \one\otimes M\otimes \one |\Phi_f\rangle\nonumber \\
    &=
    \frac{1}{2} \big{(}\langle 0|0\rangle\langle M_1\rangle\langle\phi|U_2^{\dag}U_2|\phi\rangle
    +\langle 1|1\rangle\langle M_2\rangle\langle\phi|U_1^{\dag}U_1|\phi\rangle\big{)}\nonumber\\
    &=\frac{1}{2}\left(\langle M_1\rangle + \langle M_2\rangle\right),
\end{align}
where $M_j=U_j^\dag M U_j$.

If one performs the local measurement on both target and ancilla qubits without unforking, i.e., directly on $|\Phi_2\rangle$, the measurement outcomes are
$
    \langle \Phi_2|\one\otimes M\otimes\one|\Phi_2\rangle =(\langle\psi|M_1|\psi\rangle+\langle\phi|M_1|\phi\rangle)/2
    $ and $
    \langle \Phi_2|\one\otimes \one\otimes M|\Phi_2\rangle =(\langle\psi|M_2|\psi\rangle+\langle\phi|M_2|\phi\rangle)/2.
$
Thus if $\langle\phi|M_j|\phi\rangle$ is known, the linear sum shown in equation~(\ref{eq:7}) can be calculated without the reversal c-\texttt{swap} gate at the cost of performing the local measurement on the ancilla qubit as well. We narrow our discussion to the general case without such \textit{a priori} information, and focus on the quantum forking with the unforking step.

There are several interesting remarks. First, the ancilla state $|\phi\rangle$ can be arbitrary and even unknown. As a result, one can use any state that is the easiest to prepare in the given experimental setup, such as a thermal equilibrium state or a leftover state from the previous algorithm. We assume that the computational cost of preparing such states is negligible compared to the cost of $|\psi\rangle$ preparation. Second, the same outcome can be obtained when the control qubit is initially prepared in $|1\rangle$, since this only alters the sign of the second term in equation~(\ref{eq:4}). Thus the maximally mixed state can be used as the initial state of the control qubit. This state is more difficult to prepare, but we assume that the cost is still negligible compared to that of the preparation of $|\psi\rangle$. Third, the weights $p_i$ can be manipulated by initializing the control state to either a mixed state $\sum_i p_i|i\rangle\langle i|$, or a pure state $\sum_i \sqrt{p_i}|i\rangle$. Finally, the unitary operators can be replaced with any quantum channel. The last two remarks are supported by the following. Let the initial state be represented as a density matrix $\rho_i=(p_1|0\rangle\langle 0|+p_2|1\rangle\langle 1|) \otimes \rho_\psi\otimes \rho_\phi$. The full QFS protocol produces the final density matrix $\rho_f=p_1|0\rangle \langle 0|\otimes \Lambda_1(\rho_\psi)\otimes \Lambda_2(\rho_\phi)+p_2|1\rangle \langle 1|\otimes \Lambda_2(\rho_\psi)\otimes \Lambda_1(\rho_\phi)$, where $\Lambda_i$ is a completely positive trace preserving (CPTP) map. Then the expectation value measurement gives
\begin{align}
\label{eq:8}
    \langle M\rangle = & \tr(\one\otimes M\otimes \one\rho_f)\nonumber \\
    = & p_1\text{tr}(|0\rangle \langle 0|)\text{tr}(M \Lambda_1(\rho_\psi))\text{tr}( \Lambda_2(\rho_\phi))+  p_2\text{tr}(|1\rangle \langle 1|)\text{tr}(M \Lambda_2(\rho_\psi))\text{tr}( \Lambda_1(\rho_\phi))\nonumber \\
    = & p_1\text{tr}(M \Lambda_1(\rho_\psi))+p_2\text{tr}(M \Lambda_2(\rho_\psi)).
\end{align}
Since off-diagonal terms in the density matrix of the control qubit vanish in the expectation value, equation~(\ref{eq:8}) also holds for a control qubit initialized in a pure state with probability amplitudes whose absolute squares correspond to the weights.

\subsubsection{Quadratic summation}
To evaluate the squared sum of expectation values with QFS, two qubits prepared in $|\psi\rangle$ and two arbitrary ancillae are needed. A series of c-\texttt{swap} gates initiates QF, and each pair of the qubits experiences the independent unitary evolutions denoted by $U_1$ and $U_2$, respectively. After additional c-\texttt{swap} gates unfork, the final state is given as
\begin{equation}
    |\Psi_f\rangle=\frac{1}{\sqrt{2}}\left(|0\rangle U_1|\psi\rangle U_1|\psi\rangle U_2|\phi_1\rangle U_2|\phi_2\rangle+|1\rangle U_2|\psi\rangle U_2|\psi\rangle U_1|\phi_1\rangle U_1|\phi_2\rangle\right).
\end{equation}
Now measuring $\langle M\otimes M\rangle$ on the target qubits gives
\begin{align}
   \langle M\otimes M\rangle = &\langle\Psi_f| \one\otimes M\otimes M\otimes\one\otimes\one |\Psi_f\rangle\nonumber\\
    =&\frac{1}{2}\big{(}\langle 0|0\rangle \langle \psi| M_1|\psi\rangle\langle \psi|M_1|\psi\rangle\langle \phi_1| U_2^\dag U_2|\phi_1\rangle\langle \phi_2|U_2^\dag U_2|\phi_2\rangle\nonumber \\
   &+\langle 1|1\rangle \langle \psi| M_2|\psi\rangle\langle \psi| M_2|\psi\rangle\langle \phi_1| U_1^\dag U_1|\phi_1\rangle\langle \phi_2|U_1^\dag U_1|\phi_2\rangle\nonumber\big{)} \\
   =&\frac{1}{2}\big{(} \langle M_1\rangle^2+\langle M_2\rangle^2\big{)}.
\end{align}
The procedure is shown in figure~\ref{fig:2}(b). By the same argument used above, the maximally mixed state can be used as the initial state of the control qubit, and any local CPTP map can be used instead of the local unitaries. Moreover, the control qubit can be replaced with a mixed state $\sum_i p_i|i\rangle\langle i|$, or a pure state $\sum_i \sqrt{p_i}|i\rangle$ in order to assign unequal weights. Note that more general non-linear sums, such as $\sum_i^d p_i \langle M_i\rangle\langle N_i\rangle$, can also be evaluated by measuring $\langle M\otimes N \rangle$.

\subsubsection{General summation}
In general, a density matrix in the following form can be used as an input to QFS circuit for evaluating equation~(\ref{eq:3}),
\begin{equation}
    \rho_i=\sum_{j,k}^d\sqrt{p_jp_k}|j\rangle\langle k|\otimes\rho_\psi^{\otimes q}\otimes \rho_\phi,
\end{equation}
where $\rho_\phi$ is a density matrix for $q(d-1)$ arbitrary ancilla qubits. At the end of a QFS circuit, the final density matrix is given as
\begin{equation}
\label{eq:generalfinal}
    \rho_f=\sum_{j}^dp_j|j\rangle\langle j|\otimes\Lambda_j(\rho_\psi)^{\otimes q}\otimes \rho_j+\sum_{j\neq k}^d\sum_j^d\sqrt{p_jp_k}|j\rangle\langle k|\otimes\ldots,
\end{equation}
where $\rho_j$ is the density matrix for the ancilla qubits that does not contribute to the measurement result of the target qubit. Measuring the expectation value of a $q$-local observable $\bigotimes_{j}^qM_j$ yields the weighted power sum in equation~(\ref{eq:3}).

\subsection{Projective measurement}
\label{sec:Projective}
A QFS circuit can also be followed by the projective measurement on the target qubit with an operator $\Pi_m$, which projects the target state onto the $m$th subspace with the probability $\text{Pr}(m|\rho_f)=\text{tr}\left(\Pi_m\rho_f\right)$.
Thus if $q$ copies of a target qubit undergoes independent quantum channels and are measured with a $q$-local projector, $\bigotimes_j^{q}\Pi_{j,m}$, the probability to measure the target qubit state in the $m$th subspace is
\begin{align}
    \text{Pr}(m|\rho_f)&=\sum_ip_i\text{tr}\left\lbrack \bigotimes_j^{q}\Pi_{j,m}\Lambda_i\left(\rho_\psi\right)^{\otimes q}\right\rbrack\nonumber \\
    &=\sum_ip_i\prod_j^q\text{tr}\left\lbrack \Pi_{j,m} \Lambda_i\left(\rho_\psi\right)\right\rbrack\nonumber \\
    &=\sum_ip_i\prod_j^q\text{Pr}\left\lbrack m|\Lambda_i\left(\rho_\psi\right),j\right\rbrack.
\end{align}
Therefore, the projective measurement on the target qubit yields the weighted power sum of the probabilities of an outcome, which would be obtained in a series of independent quantum measurements, without performing each measurement individually. 
\subsection{Discussion}
A general QFS for evaluating equation~(\ref{eq:3}) requires $O(\log(1/\delta)/\epsilon^2)$ experiments with $q$ target qubits, 1 control qudit of dimension $d$, $q(d-1)$ ancilla qubits in any arbitrary (even unknown) states, and $2q(d-1)$ c-\texttt{swap} gates. The control qudit can be in the mixed state where an $i$th diagonal element of the density matrix dictates the weight $p_i$. For uniform weights, i.e., $p_i=1/d\;\forall\;i$, the control qudit can be in the maximally mixed state. The advantage of QFS becomes apparent when $d$ is large and the preparation of the initial target state $|\psi\rangle$ is complex. The temporal cost of repeating the individual quantum circuits is traded for the spatial cost of having the ancilla qubits in QF. But these ancilla qubits can be in any arbitrary state, and hence the cost of preparing them is negligible. In fact, the same ancilla qubit can be repeatedly used for multiple QFS tasks without having to be reinitialized. Another notable aspect is that the dephasing noise on the control qubit does not alter the result since only the diagonal terms in the density matrix of the control qubit contribute to the measurement outcome, and the c-\texttt{swap} does not propagate phase errors from the control qubit. On the other hand, QFS may be impractical for large $q$ when the $q$-local measurement is experimentally challenging to perform.
\section{Applications}
\label{sec:app}
QFS for linear and quadratic summations can be applied to speedup various tasks in quantum science as we demonstrate with the following examples.
\subsection{Convex combination of quantum channels}
When the CPTP map $\Lambda_i$ is a unitary operator, performing the linear summation ($q=1$) using QFS naturally implements a mixed unitary channel $\Phi$, which is a convex combination of unitary channels, i.e., $\Phi(\rho)=\sum_i^d p_i U_i \rho U_i^\dag$. This can be seen in the following equation.
\begin{align}
&\text{tr}\left\lbrack\left(\one\otimes A\otimes\one^{\otimes {d-1}}\right)\rho_f\right\rbrack=\sum_i^dp_i \text{tr}\left(AU_i\rho_\psi U_i^{\dag}\right)\nonumber\\
&=\text{tr}\left(A\sum_{i=1}^dp_i U_i\rho_\psi U_i^{\dag}\right)=\text{tr}\left\lbrack A\Phi\left(\rho_\psi\right)\right\rbrack,
\end{align}
where $A$ is a Hermitian operator and $\rho_f$ is given in equation~(\ref{eq:generalfinal}) for $q=1$. Since $A$ can be any measurement operator, the expectation value measurement or the projective measurement preceded by a mixed unitary channel for an arbitrary number of unitary operators can be carried out using only a constant number of state preparations and measurements.

\begin{figure}[t]
\centering
\includegraphics[width=0.7\linewidth]{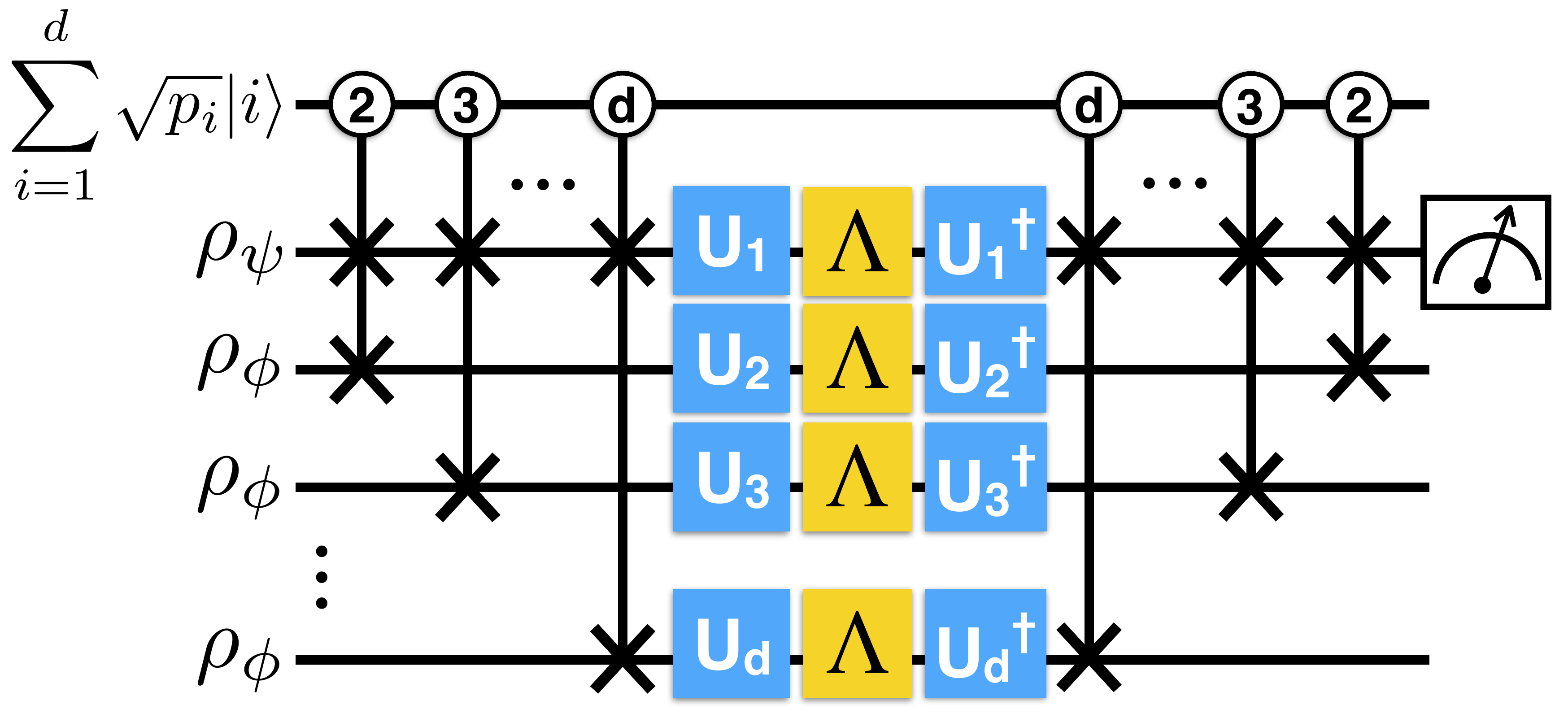}
\caption{Quantum circuit with forking for twirling a quantum channel $\Lambda$ with a finite set of unitaries $\mathcal{U}=\lbrace U_1,\ldots,U_d\rbrace$. $\rho_\psi$ represents the input density matrix that undergoes the quantum channel, and $\rho_\phi$ represents an arbitrary density matrix.}
\label{fig:3}
\end{figure}

The QFS procedure to construct a convex combination of unitary channels can naturally extend to quantum channel twirling. Twirling has been established as an important technique in quantum information science that appears in a variety of contexts, such as entanglement purification~\cite{PhysRevLett.76.722,PhysRevA.54.3824}, characterizing quantum processes~\cite{Emerson_2005,Emerson1893,PhysRevA.81.062113,PhysRevLett.106.180504,PhysRevLett.109.070504,PhysRevLett.114.140505}, studying the performance of quantum error correcting codes~\cite{PhysRevA.78.012347,PhysRevA.88.012314,PhysRevA.91.022335}, and quantum error mitigation~\cite{PhysRevX.7.021050,PhysRevX.8.031027}. Twirling a quantum channel $\Lambda$ with a finite set of unitaries, $\mathcal{U}=\lbrace U_1,\ldots,U_d\rbrace$, gives the averaged channel as
\begin{equation}
    \overline{\Lambda}(\rho)=\sum_{i=1}^dp_iU^{\dag}_i\Lambda\left(U_i \rho U_i^{\dag}\right)U_i.
\end{equation}
The quantum circuit for twirling a quantum channel $\Lambda$ with a single state preparation via quantum forking is depicted in figure~\ref{fig:3}. 
The target qubit initially given as $\rho_\psi$ and $d-1$ arbitrary ancilla qubits experience an identical quantum map $\Lambda$ in between forking and unforking steps, and the map in each qubit is conjugated by the elements of $\mathcal{U}$. The final density matrix of the QFS protocol can be written as
\begin{equation}
    \rho_f=\sum_i^dp_i|i\rangle\langle i|\otimes U_i^\dag\Lambda\left(U_i\rho_\psi U_i^{\dag}\right)U_i\otimes\rho_i+\sum_{i\neq j}^d\sum_i^d\sqrt{p_ip_j}|i\rangle\langle j|\otimes\ldots,
\end{equation}
where $\rho_i$ represents the ancilla qubits that are irrelevant to the final result of our interest. The measurement on the target qubit yields
\begin{equation}
\text{tr}\left\lbrack\left(\one\otimes A\otimes\one^{\otimes {d-1}}\right)\rho_f\right\rbrack=\text{tr}\left\lbrack A\sum_i^dp_i U_i^\dag\Lambda\left(U_i\rho_\psi U_i^{\dag}\right)U_i\right\rbrack=\text{tr}\left\lbrack A\overline{\Lambda}\left(\rho_\psi\right)\right\rbrack,
\end{equation}
where $A$ can be any measurement operator.

\subsection{Entanglemet Witness}
\begin{figure}[t]
\centering
\includegraphics[width=0.7\linewidth]{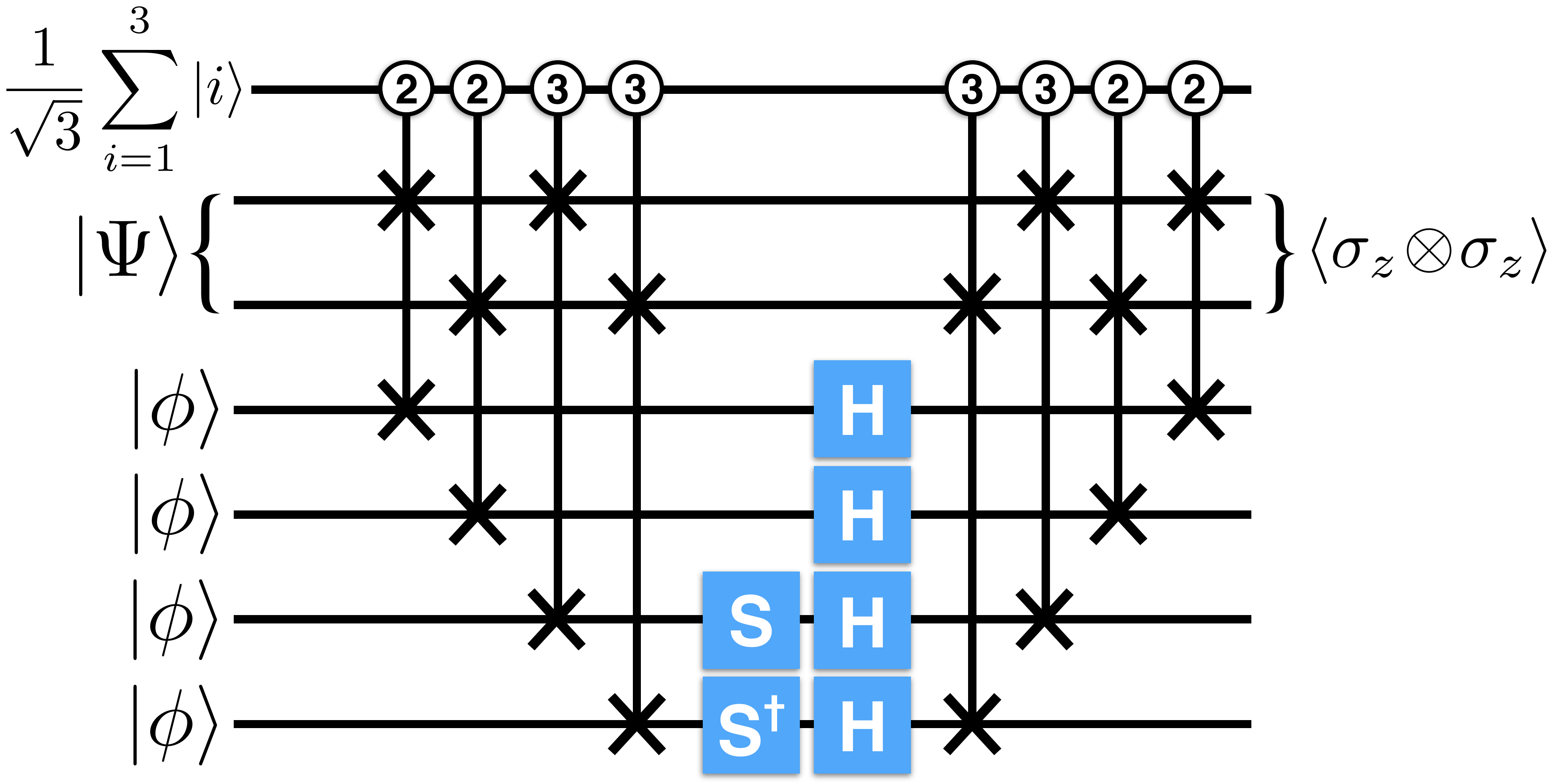}
\caption{Quantum circuit with forking for measuring entanglement witness on a bipartite state $|\Psi\rangle$ for quantum teleportation.}
\label{fig:4}
\end{figure}

The linear summation can be utilized in the experimental measurement of entanglement witnesses (EW). This also serves as an example in which QFS is useful for measuring an arbitrary observable. An EW is an observable that distinguishes an entangled state from separable ones, and is generally a useful tool in quantum information science for studying entanglement without relying on expensive quantum state tomography~\cite{1751-8121-47-48-483001}. More formally, the expectation measurement of an entanglement witness $W$ gives $\text{tr}(W\rho_{sep})\ge 0$ for all separable state $\rho_{sep}$, while there exists an entangled state $\rho_{ent}$ such that $\text{tr}(W\rho_{ent})< 0$. An EW can also be viewed as a hyperplane separating some entangled states from the set of separable states~\cite{TERHAL2000319}. Since $W$ is a Hermitian operator, it can be written as a convex combination of other Hermitian operators, i.e., $W=\sum_i c_i A_i,\;A_i=A_i^{\dagger}\;\forall i,$ and $\; c_i\in \mathbb{R}\; \forall i$. Thus, an EW can be constructed by measuring the expectation values of experimentally available observables $A_i$, and post-processing the measurement outcomes with appropriate weights. With this, the benefit of QFS becomes apparent. Instead of measuring the observables needed to construct the linear sum individually, QFS produces the same result by measuring the expectation value of a single observable, i.e., the one that is the easiest to measure in the laboratory, such as $\sigma_z$. To elucidate such a QFS application, we can consider an EW which determines whether a given state is useful for performing quantum teleportation via QFS. The teleportation witness operator can be written as $W_t=\left(\one\otimes\one-\sigma_x\otimes\sigma_x+\sigma_y\otimes\sigma_y-\sigma_z\otimes\sigma_z\right)/4$ in terms of Pauli operators~\cite{PhysRevLett.107.270501}. A naive witness experiment requires measuring three two-qubit Pauli observables. Alternatively, the same quantity can be evaluated with the QFS circuit shown in figure~\ref{fig:4}. Since there are three independent observables to be measured, i.e., $d=3$, a qutrit is needed to create three forking trajectories. Without loss of generality, the ancilla qubits are all in the same state $|\phi\rangle$ in the figure, but they can be in any state without altering the final outcome. Using $H\sigma_zH=\sigma_x$, $S\sigma_xS^\dagger=-S^\dagger\sigma_xS=\sigma_y$, where $S$ denotes the phase gate, it is straightforward to verify that the measurement of $\langle\sigma_z\otimes\sigma_z\rangle$ on the target two-qubit state yields $\langle\sigma_x\otimes\sigma_x-\sigma_y\otimes\sigma_y+\sigma_z\otimes\sigma_z\rangle$. From this, $W_t$ can be obtained. Therefore, the use of QFS reduces the number of Pauli observables to be measured from three to one.
In practice, qutrits may not be available in a given experimental setup, and it could be easier to prepare qubits. In the next example, we explain how to create three independent processes with equal weights using two qubits.
\subsection{Purity benchmarking for quantum control}
Measuring the incoherent error rate of a quantum channel has significant implications in the development of quantum devices. The incoherent error can be quantified using purity benchmarking~\cite{1367-2630-17-11-113020,PhysRevLett.117.260501}. Purity benchmarking can be combined with the standard randomized benchmarking protocol to distinguish coherent and incoherent error, which is an important step to understand different types of noise affecting quantum control. The purity benchmarking protocol requires the estimation of the purity
$
P=\sum_{j=1}^{4^n-1}\langle \mathcal{P}_j\rangle^2,
$
of a state of $n$ qubits, where $\mathcal{P}_j\in \lbrace \one,\sigma_x,\sigma_y,\sigma_z\rbrace^{\otimes n}\setminus \one^{\otimes n}$ denotes an element in the set of $n$-qubit Pauli operators minus the identity matrix. Thus, the number of expectation values to be evaluated increases exponentially with the number of qubits.
This is an example where $d$ grows large very quickly, and $q$ is small. Hence, this problem is well suited for QFS. The single qubit purity benchmarking requires the quadratic summation ($q=2$). Hence there must be two target qubits provided at the beginning of the QFS protocol, and the ability to measure two-local observables. In particular, we again assume that $\langle \sigma_z\otimes\sigma_z\rangle$ is straightforward to evaluate. Furthermore, since one needs to measure three observables, $\sigma_i$, $i\in\lbrace x,y,z\rbrace$, with equal weights, one control qutrit in the completely mixed state and four arbitrary ancillae can be used.

\begin{figure}[t]
\centering
\includegraphics[width=0.7\linewidth]{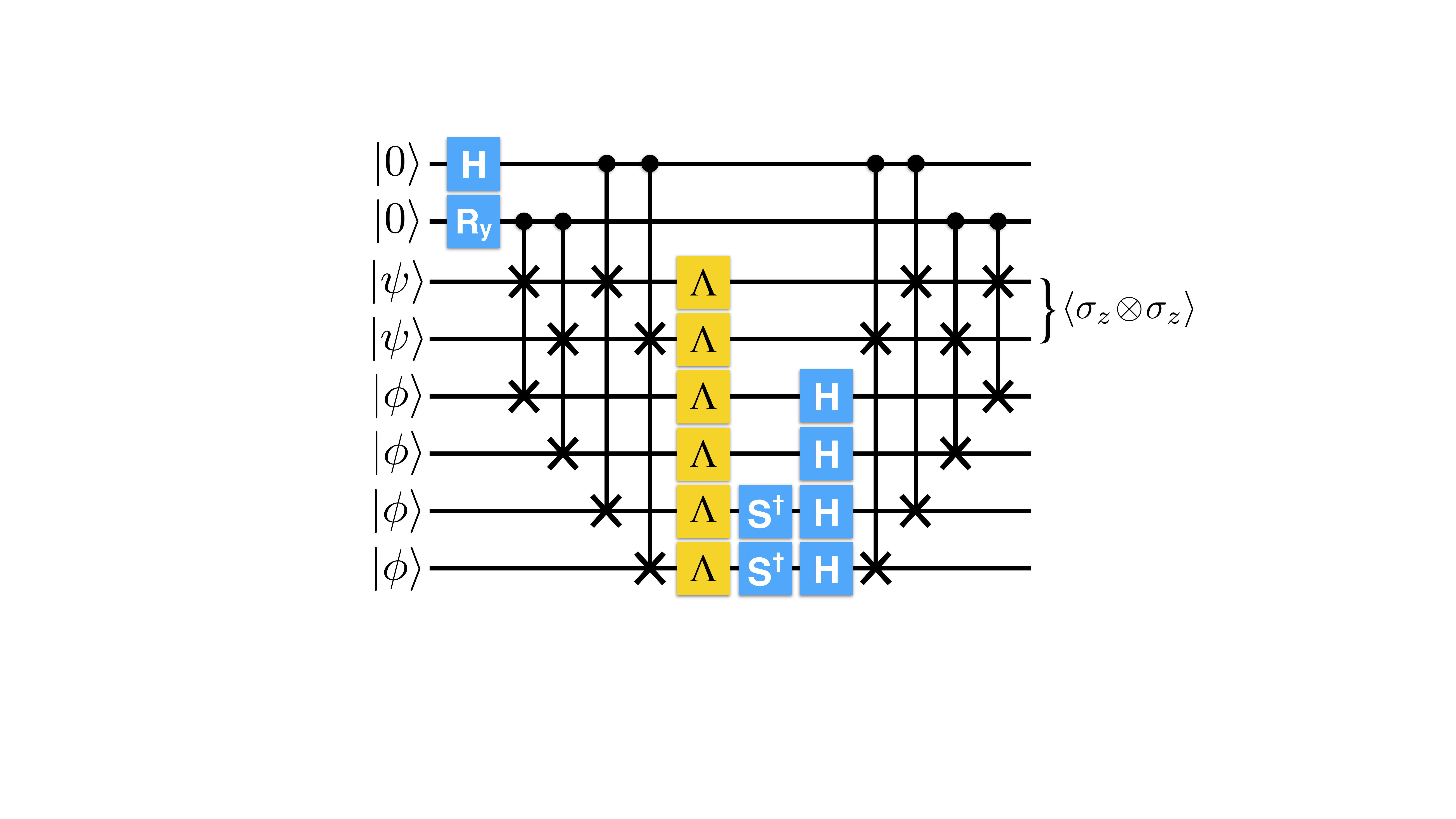}
\caption{Quantum circuit with forking for a quantum control benchmarking protocol to estimate the incoherence of noise in a quantum channel $\Lambda$. The expectation value measurement of $\sigma_z\otimes\sigma_z$ on two copies of a target qubit estimates the purity of a single qubit after being transmitted through $\Lambda$.}
\label{fig:5}
\end{figure}

In the absence of qutrits, one can use two qubits such that only three different trajectories are superposed with appropriate weights. One way to achieve this is to prepare the control qubits in
$
    H\otimes R_y(\theta)|00\rangle=
    (|00\rangle+|10\rangle)/\sqrt{3}+(|01\rangle+|11\rangle)/\sqrt{6}
$,
where $R_y(\theta)$ is the rotation around $y$-axis and $\theta=2\arccos{\sqrt{2/3}}$. Then three independent forkings can occur with the control states $|00\rangle$, $|10\rangle$, and $(|01\rangle+|11\rangle)/\sqrt{2}$, allowing the given trajectories to be measured with equal probabilities as desired. The QFS circuit for the single qubit purity measurement is depicted in figure~\ref{fig:5}. As before, the ancilla qubits are all in the same state $|\phi\rangle$, but any state can be used without altering the final outcome.
The target qubits and the ancillae first undergo the same single-qubit quantum channel $\Lambda$. But just before unforking, either $H$ or $HS^\dag $ are applied to measure $H\sigma_z H=\sigma_x$ or $S H\sigma_z HS^\dag=\sigma_y$, as desired in the purity measurement.

\section{Experiment}
\label{sec:exp}
\begin{figure}[t]
\centering
\includegraphics[width=0.8\linewidth]{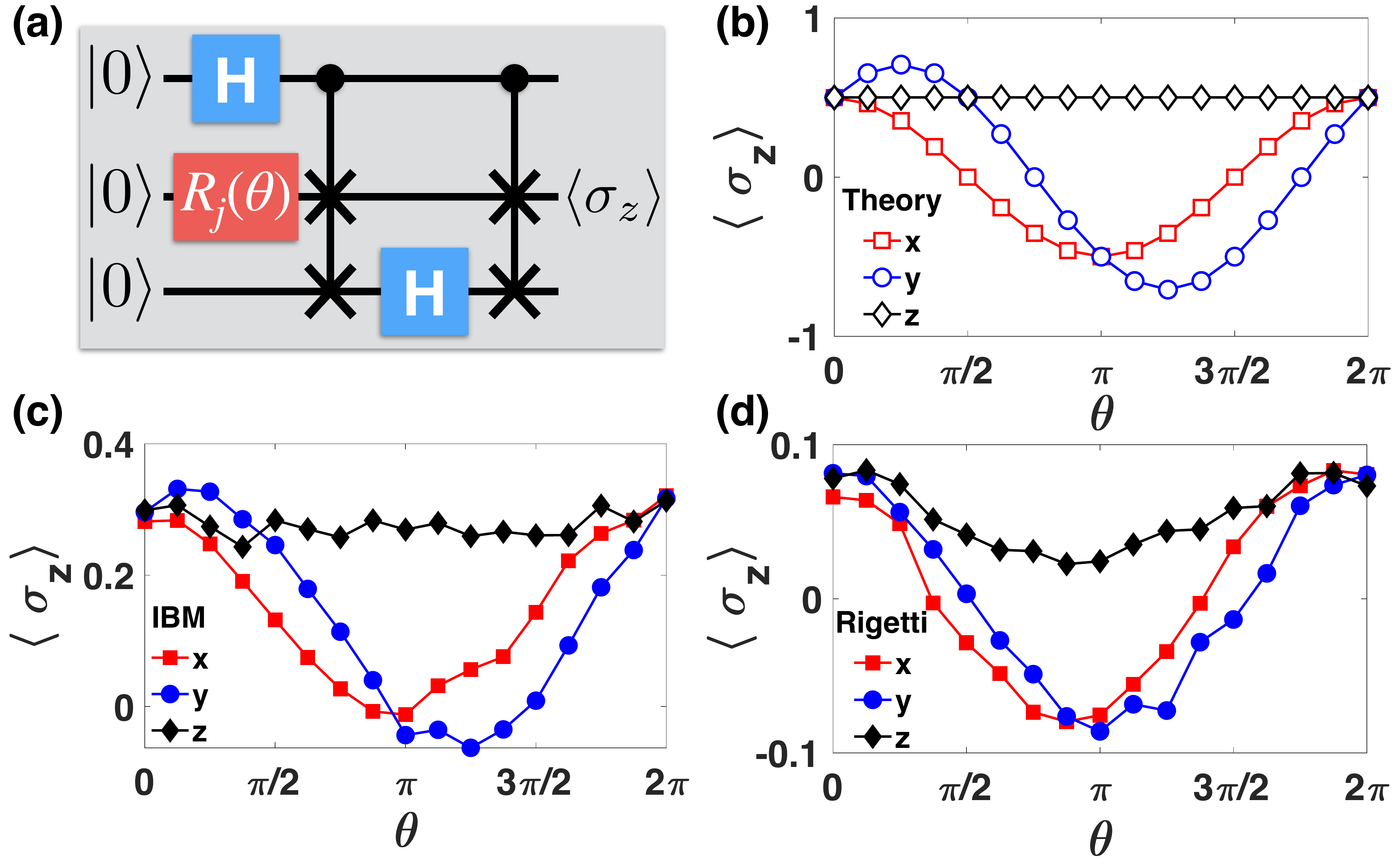}
\caption{(a) Quantum circuit with forking for discriminating the axis of the single qubit rotation in the Bloch sphere. $R_j(\theta)$ represents the single qubit rotation around $j$ by $\theta$. The $\langle\sigma_z\rangle$ measurement on the target qubit yields $\left(\langle\sigma_z\rangle+\langle\sigma_x\rangle\right)/2$, and reduces the number of Pauli observables to be measured for finding the unknown axis from two to one. Results from theory, IBM Q 5 Tenerife and Rigetti 16Q Aspen-1 are shown in (b), (c), and (d), respectively. Square, circle and diamond represent the expectation value of $\sigma_z$  obtained when  $j=x$, $y$ and $z$, respectively.}
\label{fig:6}
\end{figure}

We present the experimental results from a proof-of-principle implementation of QFS using the IBM Q 5 Tenerife~\cite{tenerife} and Rigetti 16Q Aspen-1~\cite{aspen} quantum processors. Suppose a single qubit is prepared by rotating $|0\rangle$ around an unknown axis, either x, y or z, of the Bloch sphere by a known angle $\theta$. After the state preparation, by measuring at least two Pauli observables, the axis of rotation can be discriminated. Alternatively, the sum of any two Pauli expectation values can be used. For example, $\langle\sigma_z\rangle+\langle\sigma_x\rangle$ results in $\cos(\theta)$, $\cos(\theta)+\sin(\theta)$ or $1$ for the rotation along x, y or z axis, respectively. Thus by measuring $\langle\sigma_z\rangle+\langle\sigma_x\rangle$ with respect to $\theta$ reveals the rotating axis. A simple three-qubit QFS with one Pauli expectation measurement can evaluate the same quantity. The QFS circuit implemented with IBM and Rigetti cloud quantum computers is shown in figure~\ref{fig:6}(a). After creating two forking paths via the first c-\texttt{swap} gate, a Hadamard operation is applied to the ancilla qubit for the $\langle\sigma_x\rangle$ measurement. Then the final c-\texttt{swap} gate is applied to unfork, and $\langle\sigma_z\rangle$ measurement on the target qubit yields the desired outcome.
Theoretical calculations and experimental data obtained using the IBM and Rigetti quantum processors are compared in figure~\ref{fig:6}(b), (c) and (d). The square, circle and diamond symbols represent the expectation value of $\sigma_z$ when the rotation of an angle $\theta$ is applied along the $x$, $y$ and $z$ axis, respectively. For each axis, three sets of experiments are performed with a different direction of changing $\theta$, that is varied from 0 to $2\pi$ in increments of $\pi/8$, and the results are averaged in order to suppress experimental bias in the direction of $\theta$ that may arise due to drift in calibration. In the first and second sets, $\theta$ is uniformly increased and uniformly decreased, respectively. In the last set, $\theta$ is randomly selected between 0 and $2\pi$ in $\pi/8$ increments. Each experiment is repeated for 8192 runs to collect measurement statistics. Thus each data shown in figure~\ref{fig:6}(c) and (d) is an average of 24576 runs. Despite experimental deviations from theory as illustrated with the amplitudes errors of the curves, the QFS protocol manifests successful discrimination of the initially unknown axis of the rotation with only a single expectation value measurement using currently available cloud quantum computers.

\section{Conclusion}
\label{sec:conclusion}
We developed quantum forking-based sampling as a tool to avoid redundant initial state preparations and significantly reduce the time complexity of weighted power summation, which has wide applications in quantum science. Quantum forking creates an entangled state that stores the quantum information in a different qubit in each subspace. Each subspace undergoes independent quantum trajectories in superposition. Then quantum interference enables an arbitrary weighted power summation of all outcomes from these quantum trajectories instantly in one measurement. With this technique, the $q$th power summation of $d$ measurement outcomes with arbitrary weights can be carried out with the constant cost of initial state preparation, while requiring $q(d-1)$ ancilla qubits given in arbitrary states, a control qudit with dimension $d$, $q$-local measurement and $2q(d-1)$ c-\texttt{swap} gates. The number of state preparation routines is reduced by $O(d)$. Hence QFS is particularly useful when $d$ is large, the state preparation procedure is complex, and $q$ is small. As examples, we showed how QFS can be utilized to reduce the number of state preparations and measurements in the implementation of mixed unitary channels, twirling quantum channels, entanglement witnesses and benchmarking incoherence of noise in quantum control. The proof-of-principle is demonstrated using the cloud quantum computers from IBM and Rigetti.
Our results show that for a particular family of problems, a single quantum state entangled to an arbitrary ancillary space can be exploited to provide the result as if multiple quantum states are available without violating the no-cloning theorem, paving the way for further research.
In future work, we plan to apply QFS for quantum Monte Carlo simulations, such as those used for quantum master equation unravelling, where the solutions of the quantum master equation can be obtained as an ensemble average of the solutions to the stochastic Schr\"odinger equation~\cite{TOQS}. Finding other schemes with which the quantum forking can reduce the number of state preparations, even by a constant amount, also remains an interesting open problem.

\ack{We acknowledge use of IBM Q and Rigetti Quantum Cloud Services for this work. The views expressed are those of the authors and do not reflect the official policy or position of IBM or the IBM Q team. The use of Rigetti hardware was made possible through the Developer Partnership between Rigetti and ProteinQure. We thank the Rigetti team for help associated with the Quantum Cloud Services. This work was supported in parts by the Ministry of Science and ICT, Korea, under an ITRC Program, IITP-2018-2018-0-01402, and an NRF Program, 2018K1A3A1A09078001, and by the South African Research Chair Initiative of the Department of Science and Technology and the National Research Foundation. We thank Dr. Graeme Pleasance for proofreading the manuscript.}

\bibliographystyle{iopart-num}
\bibliography{reference}

\providecommand{\newblock}{}
\begin{thebibliography}{10}
\expandafter\ifx\csname url\endcsname\relax
  \def\url#1{{\tt #1}}\fi
\expandafter\ifx\csname urlprefix\endcsname\relax\def\urlprefix{URL }\fi
\providecommand{\eprint}[2][]{\url{#2}}

\bibitem{PhysRevA.97.052329}
Sieberer L~M and Lechner W 2018 {\em Phys. Rev. A\/} {\bf 97}(5) 052329
  \urlprefix\url{https://link.aps.org/doi/10.1103/PhysRevA.97.052329}

\bibitem{PhysRevA.64.014303}
Long G~L and Sun Y 2001 {\em Phys. Rev. A\/} {\bf 64}(1) 014303
  \urlprefix\url{https://link.aps.org/doi/10.1103/PhysRevA.64.014303}

\bibitem{PhysRevA.73.012307}
Soklakov A~N and Schack R 2006 {\em Phys. Rev. A\/} {\bf 73}(1) 012307
  \urlprefix\url{https://link.aps.org/doi/10.1103/PhysRevA.73.012307}

\bibitem{Mosca:01}
Mosca M and Kaye P 2001 Quantum networks for generating arbitrary quantum
  states {\em Optical Fiber Communication Conference and International
  Conference on Quantum Information\/} (Optical Society of America) p PB28
  \urlprefix\url{http://www.osapublishing.org/abstract.cfm?URI=ICQI-2001-PB28}

\bibitem{2002quant.ph.8112G}
{Grover} L and {Rudolph} T 2002 {\em arXiv e-prints\/} quant-ph/0208112
  (\textit{Preprint} \eprint{quant-ph/0208112})

\bibitem{PhysRevA.83.032302}
Plesch M and Brukner i~c~v 2011 {\em Phys. Rev. A\/} {\bf 83}(3) 032302
  \urlprefix\url{https://link.aps.org/doi/10.1103/PhysRevA.83.032302}

\bibitem{PhysRevLett.92.177902}
Vartiainen J~J, M\"ott\"onen M and Salomaa M~M 2004 {\em Phys. Rev. Lett.\/}
  {\bf 92}(17) 177902
  \urlprefix\url{https://link.aps.org/doi/10.1103/PhysRevLett.92.177902}

\bibitem{VENTURA2000273}
Ventura D and Martinez T 2000 {\em Information Sciences\/} {\bf 124} 273--296
  ISSN 0020-0255 \urlprefix\url{https://doi.org/10.1016/S0020-0255(99)00101-2}

\bibitem{Mottonen:2005:TQS:2011670.2011675}
M\"{o}tt\"{o}nen M, Vartiainen J~J, Bergholm V and Salomaa M~M 2005 {\em
  Quantum Info. Comput.\/} {\bf 5} 467--473 ISSN 1533-7146
  \urlprefix\url{http://dl.acm.org/citation.cfm?id=2011670.2011675}

\bibitem{QRAMPhysRevLett.100.160501}
Giovannetti V, Lloyd S and Maccone L 2008 {\em Phys. Rev. Lett.\/} {\bf
  100}(16) 160501
  \urlprefix\url{https://link.aps.org/doi/10.1103/PhysRevLett.100.160501}

\bibitem{QRAMPhysRevA.78.052310}
Giovannetti V, Lloyd S and Maccone L 2008 {\em Phys. Rev. A\/} {\bf 78}(5)
  052310 \urlprefix\url{https://link.aps.org/doi/10.1103/PhysRevA.78.052310}

\bibitem{PhysRevA.86.010306}
Hong F~Y, Xiang Y, Zhu Z~Y, Jiang L~z and Wu L~n 2012 {\em Phys. Rev. A\/} {\bf
  86}(1) 010306
  \urlprefix\url{https://link.aps.org/doi/10.1103/PhysRevA.86.010306}

\bibitem{1367-2630-17-12-123010}
Arunachalam S, Gheorghiu V, Jochym-O’Connor T, Mosca M and Srinivasan P~V
  2015 {\em New Journal of Physics\/} {\bf 17} 123010
  \urlprefix\url{http://stacks.iop.org/1367-2630/17/i=12/a=123010}

\bibitem{FFQRAM-SciRep-Park}
Park D~K, Petruccione F and Rhee J~K~K 2019 {\em Scientific Reports\/} {\bf 9}
  3949 \urlprefix\url{https://doi.org/10.1038/s41598-019-40439-3}

\bibitem{UnixForking}
Ritchie D~M and Thompson K 1974 {\em Commun. ACM\/} {\bf 17} 365--375 ISSN
  0001-0782 \urlprefix\url{http://doi.acm.org/10.1145/361011.361061}

\bibitem{tenerife}
5-qubit backend: {IBM Q} team, "{IBM Q} 5 {T}enerife backend specification
  v1.3.0," (2018). Retrieved from https://ibm.biz/qiskit-tenerife. {L}ast
  {A}ccessed: 2018-11

\bibitem{aspen}
Rigetti computing, 16{Q} {A}spen-1 https://www.rigetti.com/qpu. {L}ast
  {A}ccessed: 2019-02

\bibitem{QOS}
Corrigan-Gibbs H, Wu D~J and Boneh D 2017 Quantum operating systems {\em
  Proceedings of the 16th Workshop on Hot Topics in Operating Systems\/} HotOS
  '17 (New York, NY, USA: ACM) pp 76--81 ISBN 978-1-4503-5068-6
  \urlprefix\url{http://doi.acm.org/10.1145/3102980.3102993}

\bibitem{estimate}
Huber P~J 1981 {\em Robust Statistics\/} (New York: Wiley)

\bibitem{PhysRevLett.76.722}
Bennett C~H, Brassard G, Popescu S, Schumacher B, Smolin J~A and Wootters W~K
  1996 {\em Phys. Rev. Lett.\/} {\bf 76}(5) 722--725
  \urlprefix\url{https://link.aps.org/doi/10.1103/PhysRevLett.76.722}

\bibitem{PhysRevA.54.3824}
Bennett C~H, DiVincenzo D~P, Smolin J~A and Wootters W~K 1996 {\em Phys. Rev.
  A\/} {\bf 54}(5) 3824--3851
  \urlprefix\url{https://link.aps.org/doi/10.1103/PhysRevA.54.3824}

\bibitem{Emerson_2005}
Emerson J, Alicki R and {\.{Z}}yczkowski K 2005 {\em Journal of Optics B:
  Quantum and Semiclassical Optics\/} {\bf 7} S347--S352

\bibitem{Emerson1893}
Emerson J, Silva M, Moussa O, Ryan C, Laforest M, Baugh J, Cory D~G and
  Laflamme R 2007 {\em Science\/} {\bf 317} 1893--1896 ISSN 0036-8075
  \urlprefix\url{http://science.sciencemag.org/content/317/5846/1893}

\bibitem{PhysRevA.81.062113}
L\'opez C~C, Bendersky A, Paz J~P and Cory D~G 2010 {\em Phys. Rev. A\/} {\bf
  81}(6) 062113
  \urlprefix\url{https://link.aps.org/doi/10.1103/PhysRevA.81.062113}

\bibitem{PhysRevLett.106.180504}
Magesan E, Gambetta J~M and Emerson J 2011 {\em Phys. Rev. Lett.\/} {\bf
  106}(18) 180504
  \urlprefix\url{https://link.aps.org/doi/10.1103/PhysRevLett.106.180504}

\bibitem{PhysRevLett.109.070504}
Moussa O, da~Silva M~P, Ryan C~A and Laflamme R 2012 {\em Phys. Rev. Lett.\/}
  {\bf 109}(7) 070504
  \urlprefix\url{https://link.aps.org/doi/10.1103/PhysRevLett.109.070504}

\bibitem{PhysRevLett.114.140505}
Lu D, Li H, Trottier D~A, Li J, Brodutch A, Krismanich A~P, Ghavami A,
  Dmitrienko G~I, Long G, Baugh J and Laflamme R 2015 {\em Phys. Rev. Lett.\/}
  {\bf 114}(14) 140505
  \urlprefix\url{https://link.aps.org/doi/10.1103/PhysRevLett.114.140505}

\bibitem{PhysRevA.78.012347}
Silva M, Magesan E, Kribs D~W and Emerson J 2008 {\em Phys. Rev. A\/} {\bf
  78}(1) 012347
  \urlprefix\url{https://link.aps.org/doi/10.1103/PhysRevA.78.012347}

\bibitem{PhysRevA.88.012314}
Geller M~R and Zhou Z 2013 {\em Phys. Rev. A\/} {\bf 88}(1) 012314
  \urlprefix\url{https://link.aps.org/doi/10.1103/PhysRevA.88.012314}

\bibitem{PhysRevA.91.022335}
Guti\'errez M and Brown K~R 2015 {\em Phys. Rev. A\/} {\bf 91}(2) 022335
  \urlprefix\url{https://link.aps.org/doi/10.1103/PhysRevA.91.022335}

\bibitem{PhysRevX.7.021050}
Li Y and Benjamin S~C 2017 {\em Phys. Rev. X\/} {\bf 7}(2) 021050
  \urlprefix\url{https://link.aps.org/doi/10.1103/PhysRevX.7.021050}

\bibitem{PhysRevX.8.031027}
Endo S, Benjamin S~C and Li Y 2018 {\em Phys. Rev. X\/} {\bf 8}(3) 031027
  \urlprefix\url{https://link.aps.org/doi/10.1103/PhysRevX.8.031027}

\bibitem{1751-8121-47-48-483001}
Chru{\'{s}}ci{\'{n}}ski D and Sarbicki G 2014 {\em Journal of Physics A:
  Mathematical and Theoretical\/} {\bf 47} 483001
  \urlprefix\url{http://stacks.iop.org/1751-8121/47/i=48/a=483001}

\bibitem{TERHAL2000319}
Terhal B~M 2000 {\em Physics Letters A\/} {\bf 271} 319 -- 326 ISSN 0375-9601
  \urlprefix\url{http://www.sciencedirect.com/science/article/pii/S0375960100004011}

\bibitem{PhysRevLett.107.270501}
Ganguly N, Adhikari S, Majumdar A~S and Chatterjee J 2011 {\em Phys. Rev.
  Lett.\/} {\bf 107}(27) 270501
  \urlprefix\url{https://link.aps.org/doi/10.1103/PhysRevLett.107.270501}

\bibitem{1367-2630-17-11-113020}
Wallman J, Granade C, Harper R and Flammia S~T 2015 {\em New Journal of
  Physics\/} {\bf 17} 113020
  \urlprefix\url{http://stacks.iop.org/1367-2630/17/i=11/a=113020}

\bibitem{PhysRevLett.117.260501}
Feng G, Wallman J~J, Buonacorsi B, Cho F~H, Park D~K, Xin T, Lu D, Baugh J and
  Laflamme R 2016 {\em Phys. Rev. Lett.\/} {\bf 117}(26) 260501
  \urlprefix\url{https://link.aps.org/doi/10.1103/PhysRevLett.117.260501}

\bibitem{TOQS}
Breuer H~P and Petruccione F 2007 {\em The Theory of Open Quantum Systems\/}
  (Oxford University Press) ISBN 9780199213900

\end{thebibliography}
\end{document}